\documentclass{article}
\pdfoutput=1 
\usepackage{spconf,amsmath,graphicx}
\usepackage{amssymb}
\usepackage{url}
\usepackage{balance}

\def\L{{\cal L}}
\def\inv{\vspace*{-0.15cm}}

\def\binv{\vspace*{-0.2cm}}
\title{Liver steatosis segmentation with deep learning methods}
\name{\normalsize Xiaoyuan Guo$^{\ast}$, Fusheng Wang$^{\dagger}$, George Teodoro$^{\dotplus}$, Alton B. Farris$^{\ast}$, and Jun Kong$^{\ddagger\ast}$
}

\address{
$^{\ast}$\normalsize Department of Computer Science, Emory University, Atlanta, GA, 30322, USA\\
$^{\dagger}$\normalsize Department of Computer Science, Stony Brook University, Stony Brook, NY, 11794, USA\\
$^{\dotplus}$\normalsize Department of Computer Science, University of Bras\'{\i}lia, Bras\'{\i}lia, DF, Brazil\\
$^{\ddagger}$\normalsize Department of Mathematics and Statistics, Georgia State University, Atlanta, GA, 30303, USA
}

\begin{document}

\maketitle
\begin{abstract}
Liver steatosis is known as the abnormal accumulation of lipids within cells. An accurate quantification of steatosis area within the liver histopathological microscopy images plays an important role in liver disease diagnosis and transplantation assessment. Such a quantification analysis often requires a precise steatosis segmentation that is challenging due to abundant presence of highly overlapped steatosis droplets. In this paper, a deep learning model Mask-RCNN is used to segment the steatosis droplets in clumps. Extended from Faster R-CNN, Mask-RCNN can predict object masks in addition to bounding box detection. With transfer learning, the resulting model is able to segment overlapped steatosis regions at 75.87\% by Average Precision, 60.66\% by Recall, 65.88\% by F1-score, and 76.97\% by Jaccard index, promising to support liver disease diagnosis and allograft rejection prediction in future clinical practice.
\end{abstract}

\section{Introduction}~\label{sec:intro}
Due to abnormal retention of lipids in hepatocytes, liver steatosis can result from alcohol, obesity, and type II diabetes mellitus~\cite{lee2013liver}. In addition, it serves as the hallmark of a large number of diseases, including non-alcoholic fatty liver disease (NAFLD), alcoholic fatty liver disease, and hepatotoxicity in diverse medical conditions~\cite{chalasani2012diagnosis}. Therefore, it is essential to achieve accurate quantification of steatosis droplet regions for an accurate disease diagnosis and liver transplantation evaluation~\cite{choi}. 

The prevalent gold standard for steatosis assessment is via human visual inspections of liver tissue sections, a process known to be time-consuming and subject to observer variability~\cite{Marsman2004}. Hailed as a new alternative solution, digital pathology is an emerging field that uses digital high-resolution images of tissue sections for machine-based image processing. 
Although multiple automated methods for liver steatosis measurement have demonstrated computational advantages over human reviewing process~\cite{Marsman2004,  Zaitoun, Liquori, Fiorini, roy2018segmentation, homeyer2017automated,kong2011computer}, they are not sufficiently accurate to support precise steatosis quantification, especially when overlapped steatosis regions with weak separating borders are in presence. As a result, it still remains challenging to develop a robust image analysis program to support precise liver steatosis analysis. As each tissue slide is projected to a two-dimensional microscopy image space, it is not unusual to identify a large number of tissue regions with overlapped steatosis droplets in clumps. Such spatial alignment nature, combined with substantial variations of size, staining color, and structure appearance, presents a technical barrier for individual steatosis droplet segmentation, leading to erroneous steatosis feature computation and size quantification. 

Recently, deep learning has become a successful alternative solution to biomedical image analysis. In this work, we adopt the Mask-RCNN based deep learning method~\cite{he2017mask} and successfully customize it to segment overlapped steatosis droplets in whole-slide histopathology images of live tissue sections. To establish a large training data efficiently, we propose to transfer our prior work on nuclei segmentation and have a domain expert to screen results for an accurate training data set generation~\cite{guo2018clumped}. The proposed method can separate highly clumped steatosis droplets and recover their precise contours with promising accuracy. 

\section{Method}~\label{Method}
Our work for steatosis analysis is enlightened by the Mask-RCNN segmentation method proposed for object instance segmentation~\cite{he2017mask}. Extended from Faster R-CNN~\cite{ren2015faster}, Mask-RCNN replaces the Region of Interest (ROI)-pooling operation with `ROI-Align' for solving the misalignment problem. This change in architecture enables segmentation of individual objects from different categories and results in substantial improvement in the segmentation accuracy. Due to the promising performance of Mask-RCNN for instance segmentation, we propose to customize this architecture for steatosis segmentation. 
\begin{figure}[htb]
\centering
\includegraphics[width=8.5cm, height=9cm]{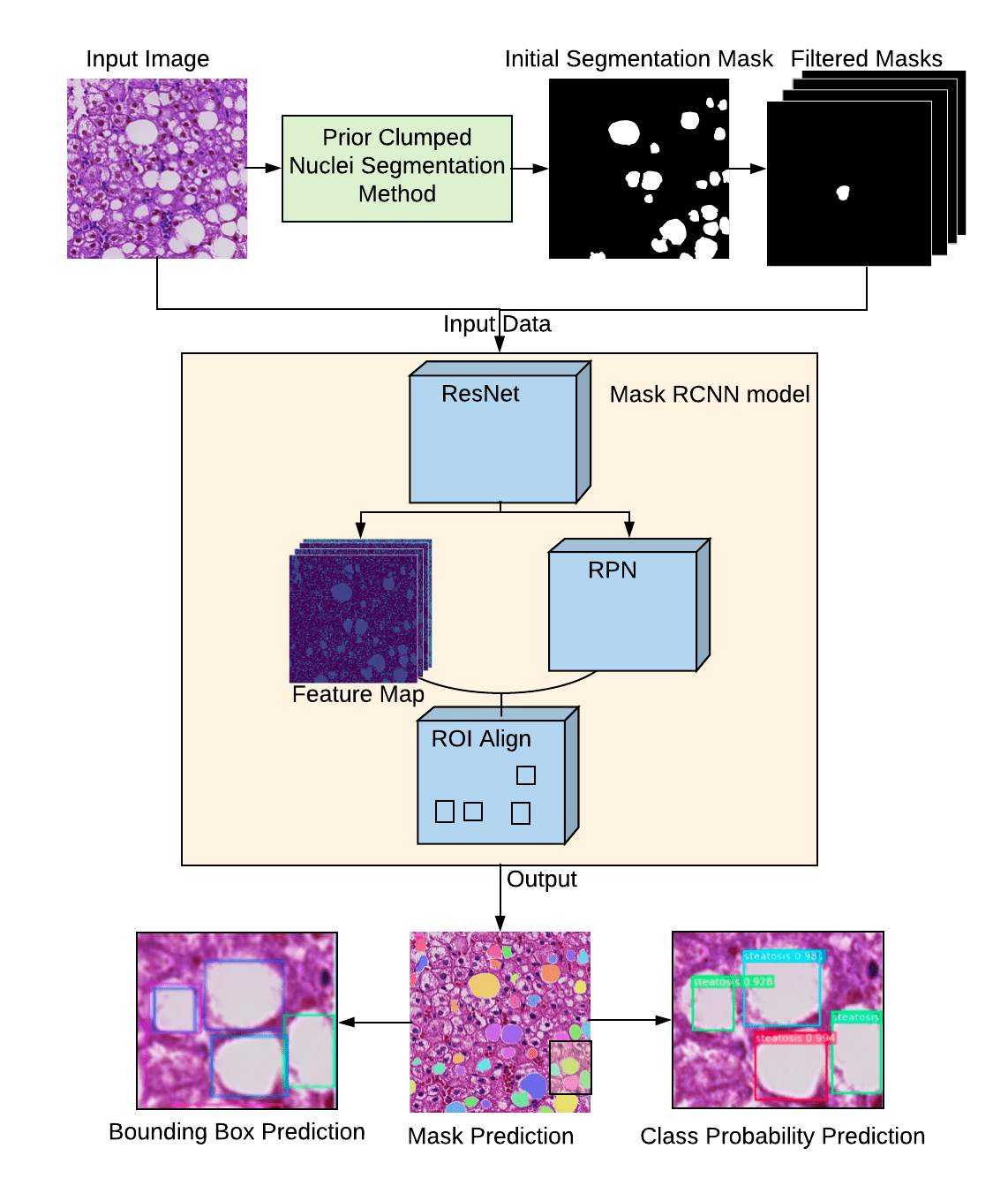}\inv\inv\inv
\caption{Schema of steatosis segmentation method.}
\label{fig:procedure}\inv\inv\inv
\end{figure}
The method schema is presented in Fig.~\ref{fig:procedure} where three primary components are presented: training data preparation with our prior work on nuclei segmentation~\cite{guo2018clumped}, model training with transfer learning, and overlapped steatosis segmentation in testing images. 

\inv
\subsection{Training Data Preparation}
Due to the two-dimensional image space projection, a large amount of densely aligned steatosis droplets can touch together with blurred dividing boundaries. Therefore, it is not feasible for pathologists to annotate all steatosis masks for efficient training data production. This is a common problem for deep learning model training in a wide scope of research investigations.

To facilitate training data preparation, we modify our previous nuclei segmentation method~\cite{guo2018clumped} and generate initial segmentation masks for steatosis instances. As our previous method aims at separating clumped nuclei in fluorescence in-situ hybridization images, it can not be directly applied to bright field histopathology images for steatosis droplet segmentation. As a result, we modify such method as follows. First, we convert the input color image to its gray-scale representation that is further binarized by a normalized threshold. All non-tissue areas in the image background are excluded for further analysis. 
Next, overlapped steatosis candidates are identified by rejecting connected foreground regions where solidity is over 0.95. A high curvature point voting method is used to detect dividing candidate points. They are connected based on the fitting quality evaluation by an ellipse fitting model, spatial proximity, shape convexity, and curvature information. Finally, we recover dividing curves by local shape based intensity analysis in a sector-shaped searching space, and produce the corresponding isolated steatosis masks~\cite{guo2018clumped}.

Although this process produces satisfied results for a large number of touching steatosis droplets, there are partitioned steatosis instances that are not matched with their histology structures as reviewed by the domain expert. 
Such results are removed from the training data set. 
As the number of such instances is limited, these unlabeled foreground regions would have little impact on the generalization ability of the trained model. In this way, a good training data set is established in an efficient manner. The resulting data set includes 451 images of $1024\times1024\times3$ with corresponding mask sets. Each image patch $I$ has multiple masks $\left\{ { {M}_{1} },{ {M}_{2},{ {M}_{3},\cdots, { {M}_{n} } } } \right\}$, with each mask image containing one steatosis droplet, essential for solving the overlapped steatosis problem. With this training data set, Mask-RCNN is able to learn how to segregate overlapped steatosis droplets through image-mask pairs. 

\inv
\subsection{Deep Learning Model}
There are three primary components in Mask-RCNN: the backbone, Region Proposal Network(RPN), and `ROI-Align'. The backbone is composed with Convolutional Neural Networks (CNNs) that can extract multi-level image features. We use modified resnet41, resnet50, and resnet65~\cite{he2016deep} as our backbone CNNs. The second component RPN scans the input image with a sliding-window and detects steatosis droplet regions in our study. `ROI-Align' further analyzes ROIs from the RPN and interpolates the feature maps from the neural network backbone at multiple locations. In this way, it can handle the incorrect alignment from Faster R-CNN~\cite{ren2015faster}. With these deep learning components, the resulting model can classify objects into different classes, provide object positions with bounding boxes, and produce a mask for each detected object. In the training process, these three components are orchestrated to minimize the following multi-mask loss function for each steatosis instance~\cite{he2017mask}:
 \begin{equation}
 \L = \L_{cls} + \L_{bbx} + \L_{mask} 
 \end{equation}
 \noindent 
where $\L_{cls}, \L_{bbx}$, and $\L_{mask}$ are classification loss, bounding box loss, and mask prediction loss, respectively. More specifically, $\L_{cls}=-\log(p_i)$ where $p_i$ is the class probability of the instance $i$; 
$\L_{bbx}=\sum_{c_i(j) \in \{ x_i,y_i,w_i,h_i\} } SmoothL_1(c_i(j) - C_i(j))$, where $c_i$ and $C_i$ are the centroid coordinates, width, and height of predicted and ground-truth bounding box for the instance $i$. Additionally, we have the function $SmoothL_1(\mathord{\cdot})$ defined as: 
$SmoothL_1(x) =\left\{\begin{matrix}0.5{ x }^{ 2 },\quad \quad \quad \mathrm{if}  \left| x \right|<1 \\ \left| x \right|-0.5,\quad \quad \mathrm{otherwise} \end{matrix} \right.$.

\noindent$\L_{mask}=-\frac{ 1 }{N^2} \sum\limits_{ 1\le i,j\le n }  [P_{ ij }\log { { p }_{ ij } } +(1-P_{ ij })\log  (1-{ p }_{ ij })] $, where ${p}_{ij}$ is the predicted mask probability and $P_{ij}$ is the ground-truth mask label at pixel $(i, j)$ in a $N\times N$ region. 

Typical steatosis segmentation results are demonstrated at the bottom part of Fig.~\ref{fig:procedure}. The output image on the left presents the steatosis mask prediction with individual steatosis objects color-coded, whereas output images in the middle and right illustrate the predicted steatosis bounding boxes and the classification probabilities, respectively. 

\begin{figure}[b]
\includegraphics[width=0.5\textwidth, height=2.7cm]{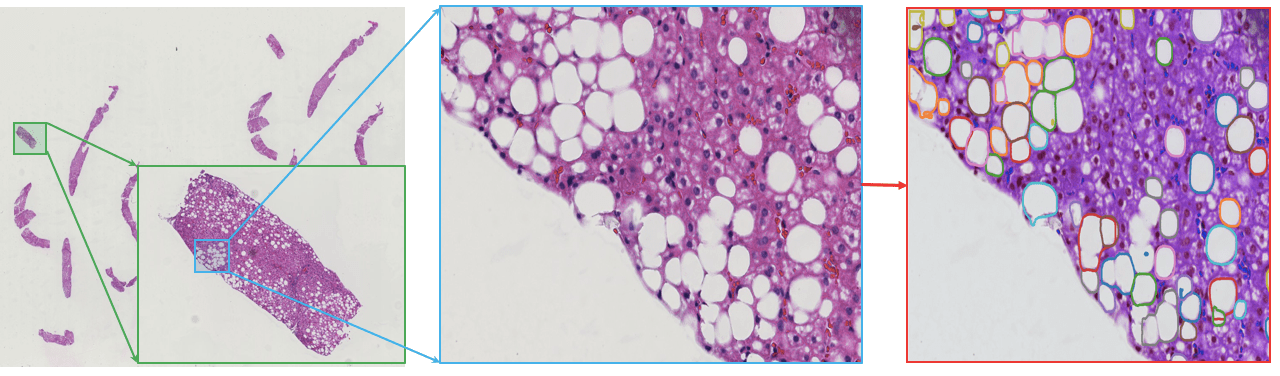}\inv\inv\inv
\caption{Segmentation process for a whole-slide microscopy image.}
\label{fig:process}\inv\inv\inv
\end{figure}

\section{Experiments}~\label{sec:experiment}
With our prior segmentation method for nuclei~\cite{guo2018clumped}, we generate segmentation masks at the highest image resolution. Our training data set is efficiently generated after a domain pathologist removes erroneous masks. The final data set contains 451 liver images of $1024\times 1024\times 3$ with ground-truth masks of which 387, 45, and 19 images are used for training, validation, and testing, respectively.

Random neural network initializations can result in an overwhelmingly expensive time cost for model training. Demonstrating its strength for problem solving at a reduced computational cost, transfer learning~\cite{pan2010survey} enables the pre-trained models to serve as the initial point for customized training, and has become popular in a large number of deep learning studies. In our experiment, network weights from the pre-trained COCO model are adopted to initialize our training process. Three backbones network structures, i.e. modified Resnet41, Resnet50 and Resnet65, are used, respectively. For these network backbones, we train the head layer for 30, 20, and 30 epochs, respectively. After head training, all layers are trained to achieve the best segmentation accuracy with 80, 50, 50 epochs, respectively. These epoch numbers are determined heuristically. To minimize the total loss in the training process, back-propagation and Stochastic Gradient Descent(SGD) are utilized. We run experiments on two GPUs (12GB RAM Tesla K80, NVIDIA Inc.) for 300 iterations, with six images per GPU for each mini-batch. The initial learning rate is 0.02, decreased by 10-fold for each 300 iterations. Additionally, online data augmentation techniques are used to further scale up data set size, improve the training performance, and increase generalizability and robustness of the trained model. These techniques include random affine transform, random flipping, and Gaussian blurring. 

The process of applying the trained network to whole-slide microscopy images is presented in Fig.~\ref{fig:process} where the left image presents the overall view of a representative whole-slide image, with a green box illustrating the close-up view of a small tissue part. The image in the middle demonstrates clustered steatosis droplets in a small liver tissue region at the full image resolution, while segmentation results of steatosis droplets by our trained network are illustrated with distinct colors in the right image.

\begin{figure}[tb!]
\centering
\includegraphics[width=9cm, height=5cm]{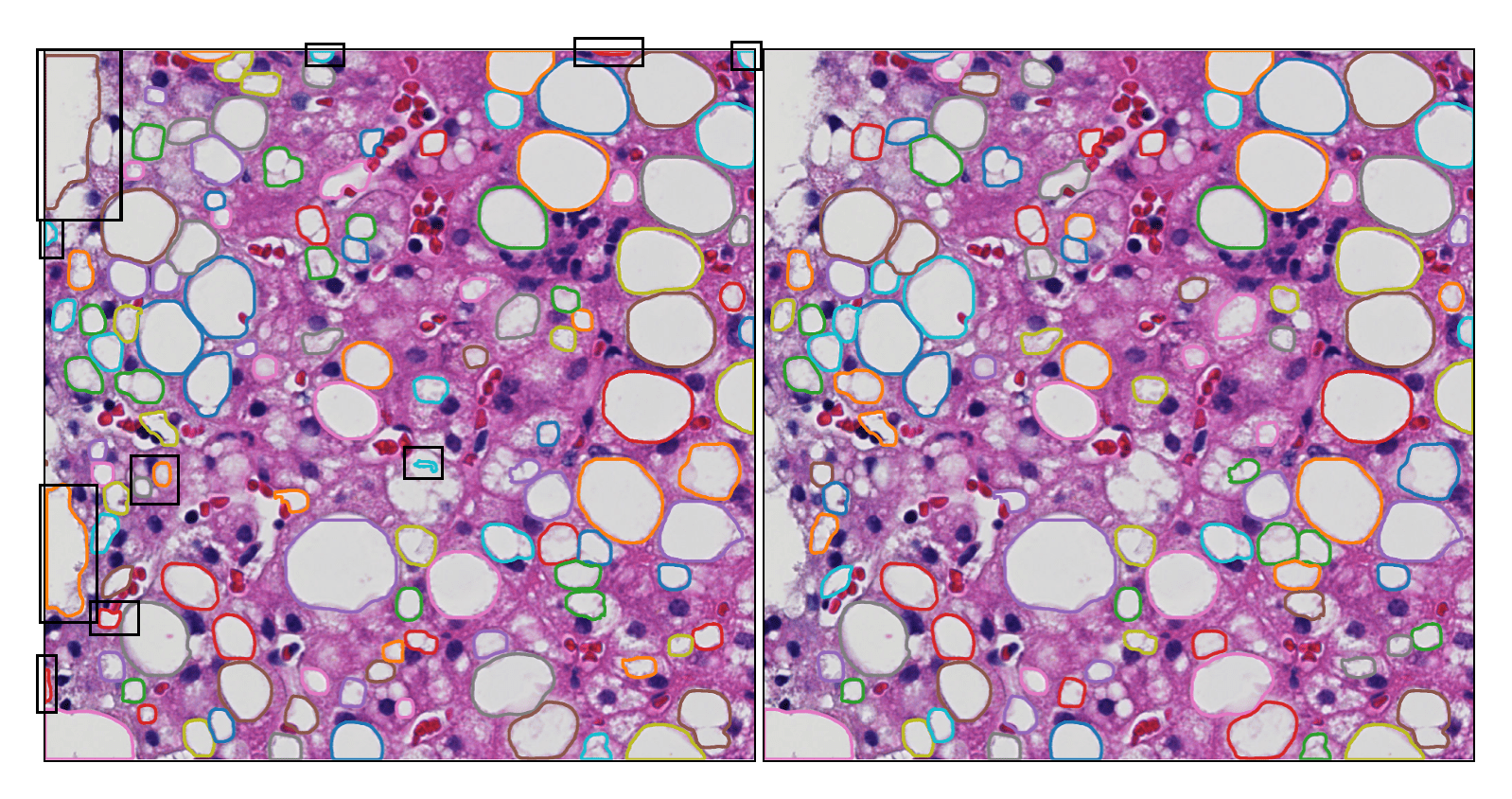}\inv\inv\inv
\caption{Segmentation result (Left) before and (Right) after post-processing, with black boxes in the left image indicating the discarded steatosis regions.}
\label{fig:filter}\inv
\end{figure}

To make this method flexible for diverse liver disease diagnosis and transplantation evaluation settings, we characterize each recognized steatosis candidate by the eccentricity, size, and perimeter, and record the segmentation score resulting from the neural network prediction. We provide user-defined thresholds for these features, enabling customized steatosis droplets retention.
In this way, domain pathologists can have a convenient way to select cutoff values for these features and get readily informed of the number and the morphological profiles of the retained steatosis objects. We demonstrate the neural network segmentation results before and after such post-processing in Fig.~\ref{fig:filter} where 
we keep steatosis candidates with size, perimeter and eccentricity within $[0.001, 6]$, $[0.5, 4]$, and  $[0.2, 1.5]$ times of the average steatosis, respectively.

\begin{figure*}[tb]
\includegraphics[width=1.0\textwidth]{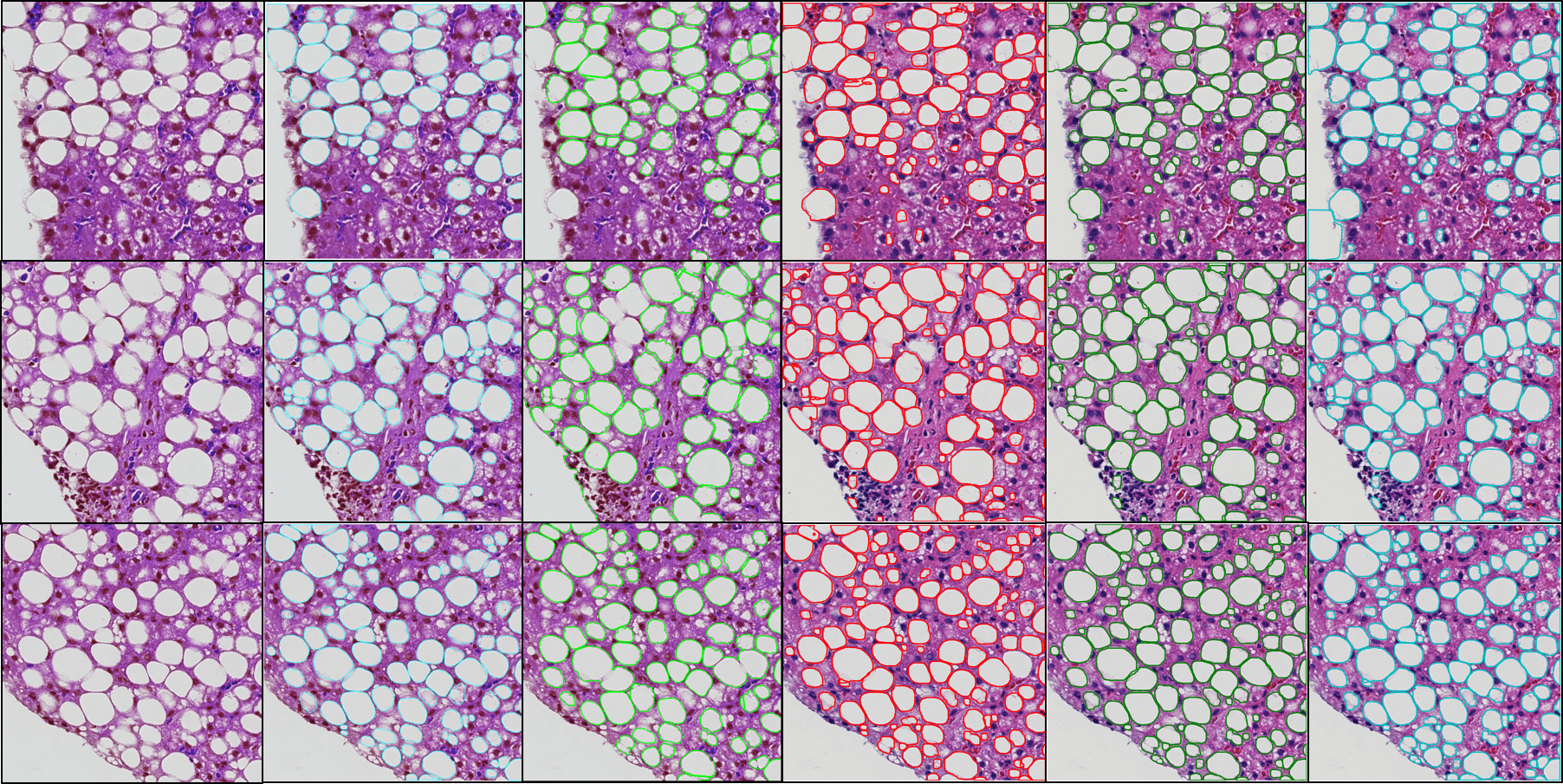}\inv\inv
\caption{In columns from left to right, we demonstrate original images, ground-truth, segmentation results from method~\cite{guo2018clumped}, Mask-RCNN with modified Resnet41, Resnet50, and modified Resnet65, respectively.}
\label{fig:res}\inv
\end{figure*}

Representative segmentation results are presented in Fig.~\ref{fig:res} where the original input images, ground-truth segmentation, results from our earlier method~\cite{guo2018clumped}, outputs of presented deep learning approach with modified Resnet41, Resnet50 and modified Resnet65 backbones are presented in columns from left to right. It is noted that deep learning methods, especially the network with backbone Resnet50, present better results than our earlier work as they present less under-segmentation results. Additionally, deep learning methods present good performances on handling clumped steatosis clusters with complex topology. 

To quantitatively evaluate the presented method, we compare results from deep learning algorithm with the ground-truth data. Table~\ref{tab:1} presents evaluation results of averaged steatosis measures by four metrics, including Average Precision, Recall Ratio, F1-score, and Jaccard index. It is noticed that the trained network with Resnet50 achieves the best precision, recall ratio, F1-score, and Jaccard index. 

\inv
\section{Conclusion}\inv
We propose to use the deep learning method to solve the overlapped liver steatosis segmentation problem. Due to lack of labelled steatosis droplets from liver microscopy images for training, we modify our earlier nuclei segmentation method to generate liver steatosis training data after an efficient screening process by a domain expert. The trained neural network model is demonstrated to segment liver steatosis droplets, especially those in clumps with promising accuracy. Quantitative evaluations suggest that deep learning technology enables accurate and high-performance steatosis segmentation, a promising tool for enhancing liver disease diagnosis and transplantation assessment.

\begin{table}[tb]
\caption{Method performance evaluation and comparison.}
\begin{tabular}{|c|c|c|c|c|c|}\hline
Method & AP & Recall & F1-score& Jaccard\\ \hline
Clump\_seg~\cite{guo2018clumped} & 52.18\%  & 45.50\% & 45.03\% & 67.42\%\\ \hline
ResNet41 & 67.61\%  &58.37\% & 62.06\%& 73.18\% \\ \hline
ResNet50& \textbf{75.87\%}  & \textbf{60.66\%} & \textbf{65.88\%} &\textbf{76.97}\%\\ \hline
ResNet65 & 69.55\%  &55.60\% & 61.69\%& 74.38\% \\ \hline
\end{tabular}
\label{tab:1}
\end{table}


\inv
\section{Acknowledgement}\inv
This research is supported in part by grants from National Institute of Health K25CA181503 and National Science Foundation ACI 1443054 and IIS 1350885, and CNPq.

\balance

\inv
\bibliographystyle{IEEEbib}

\end{document}